\documentclass[a4paper,aps,prd,10pt,preprintnumbers,twocolumn,superscriptaddress,nofootinbib,amsmath,amssymb]{revtex4-1}
\usepackage{graphicx}
\usepackage[utf8]{inputenc}
\usepackage[T1]{fontenc}
\usepackage{cmap}
\def\imo{i}

\def\K{{\cal K}}
\DeclareMathAlphabet{\pazocal}{OMS}{zplm}{m}{n}

\begin{document}
\title{Quasinormal modes and Hawking radiation of black holes in cubic gravity}
\author{R. A. Konoplya} \email{roman.konoplya@gmail.com}
\affiliation{Research Centre for Relativistic Physics and Astrophysics, Institute of Physics, Silesian University in Opava, CZ-746 01 Opava, Czech Republic}
\affiliation{Peoples Friendship University of Russia (RUDN University), 6 Miklukho-Maklaya Street, Moscow 117198, Russian Federation}
\author{A. F. Zinhailo} \email{antonina.zinhailo@physics.slu.cz}
\affiliation{Research Centre for Relativistic Physics and Astrophysics, Institute of Physics, Silesian University in Opava, CZ-746 01 Opava, Czech Republic}
\author{Z. Stuchlik} \email{zdenek.stuchlik@physics.slu.cz}
\affiliation{Research Centre for Relativistic Physics and Astrophysics, Institute of Physics, Silesian University in Opava, CZ-746 01 Opava, Czech Republic}
\begin{abstract}
We consider quasinormal modes and Hawking radiation of four-dimensional asymptotically flat black holes in the most general up to-cubic-order-in-curvature dimension-independent Einsteinian theory of gravity that shares its graviton spectrum with the Einstein theory on constant curvature backgrounds. We show that damping rate and real oscillation frequencies of quasinormal modes for scalar, electromagnetic and Dirac fields are suppressed once the coupling with the cubic term is on. The intensity of Hawking radiation is suppressed as well, leading to, roughly, one order longer lifetime at a sufficiently large coupling constant.
\end{abstract}
\maketitle

\section{Introduction}

A  theory of gravity has been proposed in \cite{Bueno:2016xff}  which is the most general up to-cubic-order-in-curvature theory of gravity that shares its graviton spectrum with Einstein theory on a constant curvature background. This Einsteinian cubic theory of gravity is not trivial in four dimensions and therefore recently it has attracted considerable interest \cite{Quiros:2020uhr,Marciu:2020ysf,KordZangeneh:2020qeg,Burger:2019wkq,Jiang:2019kks,Emond:2019crr,Erices:2019mkd,Mehdizadeh:2019qvc,Li:2019auk,Arciniega:2018fxj,Bueno:2018xqc,Bueno:2017sui,Hennigar:2017ego,Bueno:2017qce}. The numerical solution representing the asymptotically flat black hole in this theory was obtained in \cite{Hennigar:2016gkm,Bueno:2016lrh}, and  the analytical approximation of the black hole metric was obtained in \cite{Hennigar:2018hza} using the general parametrization for spherically symmetric metrics suggested in \cite{Rezzolla:2014mua}. Further properties of this black hole, such as gravitational lensing and particle motion were studied in \cite{Poshteh:2018wqy,Hennigar:2018hza}. Theories with higher curvature corrections form an important class of theories which also appear in the low-energy limit of string theory, and therefore, black holes were extensively investigated in such theories of gravity (see, for example, \cite{Kanti:1995vq} and references therein).

One of the most important characteristics of black hole geometry is its quasinormal spectrum \cite{Konoplya:2011qq}. Quasinormal modes dominate in the late time (ringdown) phase of the black hole's response to external perturbations. They are currently observed when detecting gravitational wave from astrophysical black holes \cite{Abbott:2016blz,TheLIGOScientific:2016src}. At the same time the current uncertainty in measurements of mass and angular momentum of black holes leaves considerable room for alternative theories of gravity \cite{alternative}, and the study of quasinormal spectra of black holes in various alternative theories of gravity is a necessary tool for further constraining of these theories.

Another characteristic, essential for primordial and sufficiently small black holes, is Hawking radiation in the vicinity of the black hole horizon \cite{Hawking:1974sw}. Higher curvature corrections could represent quantum corrections to the black hole geometry and is, therefore, important in the regime of intensive Hawking evaporation. As it was shown for black holes with quadratic corrections in curvature, Hawking radiation is considerably affected by higher curvature corrections \cite{Konoplya:2020cbv,Konoplya:2019ppy,Zhang:2020qam,Li:2019bwg}, even when the deformation of the geometry is relatively small \cite{Konoplya:2010vz,Konoplya:2019hml}. In particular for higher dimensional Einstein-Gauss-Bonnet black holes \cite{Rizzo,Konoplya:2010vz} intensity of Hawking radiation of a black hole whose spacetime is only slightly deformed from the Tangherlini geometry may differ by a few orders.  Therefore, it is tempting to learn whether the intensity of Hawking radiation is so sensitive characteristic in the Einsteinian cubic gravity as well.

Finally, analysis of various radiation phenomena for the analytical approximation of the numerical black hole solution obtained in \cite{Hennigar:2018hza} at different orders of this approximation is interesting, because it allows us to test the accuracy of the analytical approximation in the context of the recent statement that spherically symmetric and asymptotically flat black holes can very well be described by only three parameters within this parametrization \cite{Konoplya:2020hyk}. Thus, looking at quasinormal modes of the above black hole with cubic curvature corrections when the metric is represented with various order of accuracy, that is, with larger or smaller number of parameters,  we can have another test of this statement \cite{Konoplya:2020hyk}.

Having all the above motivations in mind we will study quasinormal modes of scalar, electromagnetic and Dirac fields in the background of the four-dimensional spherically symmetric and asymptotically flat black hole in the Einsteinian cubic theory of gravity. We will also calculate grey-body factors of test fields for this case, and estimate the intensity of Hawking radiation. It will be shown that both real and imaginary part of quasinormal modes, representing respectively the real oscillation frequency and damping rate, are suppressed due to the cubic corrections. The intensity of Hawking radiation is also considerably decreased by the cubic corrections.

The paper is organized as follows. In Sec. II we summarize the basic information on the Einsteinian cubic gravity and analytical approximation for the black hole metric obtained in \cite{Hennigar:2018hza}. Section III is devoted to calculations of quasinormal modes. In Sec. IV we calculate grey-body factors for test fields, while in Sec. V we find the energy emission rate and lifetime of the black hole under consideration. Finally, in the Discussion we summarize the obtained results and discuss open problems.

\section{The black hole metric}
The action for the Eisnteinian cubic gravity (ECG) has the form \cite{Bueno:2016xff},
\begin{equation}
S=\frac{1}{16 \pi}\int \! d^{4}x \, \sqrt{-g} \left[R-\frac{\lambda}{6} \mathcal{P} \right],
\end{equation}
where $R$ is the usual Ricci scalar and
\begin{align}
\mathcal{P} =& \, 12 R_a{}^b{}_c{}^d R_b{}^e{}_d{}^f R_e{}^a{}_f{}^c + R_{ab}^{cd}R_{cd}^{ef}R_{ef}^{ab}
\nonumber\\
&- 12 R_{abcd}R^{ac}R^{bd} + 8 R_a^b R_b^c R_c^a \, .
\end{align}
Here $\lambda$ is the coupling constant, representing ``the weight'' of the cubic term.

Static, spherically symmetric solution was numerically obtained  in  \cite{Hennigar:2016gkm,Bueno:2016lrh} and has the following form:
\begin{equation}
ds^2 = -f dt^2+\frac{1}{f}dr^2+r^2d\Omega_{(2)}^2.
\end{equation}
The field equation for the metric function $f(r)$ is:
\begin{align}
&
2M = -(f-1) r-\lambda [\frac{f'^{3}}{3}+\frac{f'^{2}}{r}-\frac{2}{r^2} f(f-1)f'
\\
&\hspace{1.2cm}
-\frac{1}{r} ff'' (r f'-2(f-1))].
\nonumber
\end{align}
The mass and Hawking temperature of the black hole are given by the following relations  \cite{Hennigar:2016gkm,Bueno:2016lrh}:
\begin{subequations}
\begin{eqnarray}
M=\frac{r_0^3}{12 {\lambda}^2} \left[r_0^6+(2 \lambda-r_0^4) \sqrt{r_0^4+4 \lambda}\right],\\
T=\frac{r_0}{8 \pi \lambda} \left[\sqrt{r_0^4+4 \lambda}-r_0^2\right],
\end{eqnarray}
\end{subequations}
where $r_{0}$ is the radius of the event horizon.
Following \cite{Rezzolla:2014mua} the metric function can be represented in the following form \cite{Hennigar:2018hza}:
$$f(x) = $$
\begin{equation}
x\left[1-\varepsilon (1-x)+(b_0-\varepsilon)(1-x)^2+\widetilde{B}(x)(1-x)^3\right],
\end{equation}
where $x$ is a new compact coordinate,
\begin{equation}
x =1-\frac{r_0}{r},
\end{equation}
and
\begin{equation}
\widetilde{B}(x) =\frac{b_1}{1+\frac{b_2 x}{1+\frac{b_3 x}{1+\cdots}}}.
\end{equation}
The above expressions represent an approximation of the numerical metric function in the whole space from the event horizon to infinity. This kind of representation was used to approximate numerical black hole solutions in a number of other theories, for example, in the Einstein-dilaton-Gauss-Bonnet \cite{Kokkotas:2017ymc}, Einstein-scalar-Gauss-Bonnet \cite{Konoplya:2019fpy}, Einstein-Weyl \cite{Kokkotas:2017zwt} and scalar-Maxwell \cite{Konoplya:2019goy}, quartic \cite{Khodabakhshi:2020hny} theories of gravity. This parametrization has been also extended to the axially symmetric spacetimes \cite{Konoplya:2016jvv,Younsi:2016azx}, representing rotating black holes.  The privilege to use this continued fraction expansion is the superior convergence of the expansion which usually provides a compact analytical form approximating the numerical metric with sufficient accuracy.

\begin{figure}
\resizebox{\linewidth}{!}{\includegraphics*{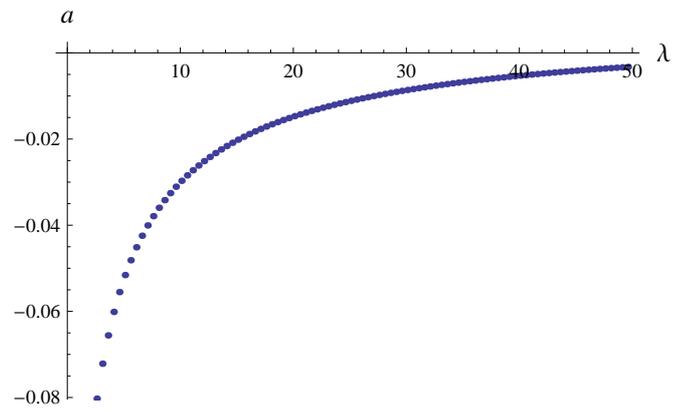}}
\caption{The parameter $a$ as a function of the coupling constant $\lambda$ in the units $M=1$.}\label{fig:a}
\end{figure}

Parameter $\varepsilon$ determines the deviation of the radius of the event horizon from the Schwarzschild radius:
\begin{equation}
\varepsilon=\frac{2 M}{r_0}-1,
\end{equation}
while in order to match current values of post-Newtonian parameters, one must have
\begin{equation}
b_0 =0.
\end{equation}
The remaining coefficients $b_1$, $b_2$  etc. are fixed by the behavior of the metric near the event horizon and can be expressed in terms of $T$, $M$ as follows:
\begin{equation}
b_1 =4 \pi r_0 T+\frac{4 M}{r_0}-3,
\end{equation}
\begin{equation}
b_2 =-\frac{r_0^3 a +16 \pi r_0^2 T+6 (M-r_0)}{4 \pi r_0^2 T+4 M-3 r_0}.
\end{equation}
Here, for small and moderate values of the coupling constant $\lambda$, the coefficient $a$ can be approximated by Eq. 16 of \cite{Hennigar:2018hza}, while in the general case it can be found only numerically, and here we plot the values of the parameter $a$ as a function of $\lambda$ (see Fig. \ref{fig:a}). Higher order correction is given by the nonzero parameter $b_3$, the explicit form for which can be found in the Appendix of \cite{Hennigar:2018hza}. However, as we will see that even the first order expansion given by the nonzero $b_1$ is sufficiently accurate, so that the second coefficient $b_2$ only slightly correct the observable quantities at sufficiently large values of the coupling constant $\lambda$. Thus, there is no practical sense to use the third order expansion for the metric.  With the above equations at hand we can analyze quasinormal modes and Hawking radiation for this black hole metric.

\section{Quasinormal modes of scalar, Dirac and electromagnetic fields}

In this section we will study quasinormal modes of a scalar, Dirac and electromagnetic fields. The reduction of the perturbation equation to the master wavelike form for gravitational perturbations is highly nontrivial problem for the above theory and deserves separate consideration. However, in a plenty of cases the behavior of the quasinormal spectrum for test and gravitational fields is qualitatively the same and approaches the universal regime, independent of the spin of the field in the high frequency (eikonal) limit. The eikonal quasinormal modes of test fields are known to be dual to some characteristics of null geodesics \cite{Cardoso:2008bp,Konoplya:2017wot}. Moreover, already at sufficiently small values of $\ell$ the quasinormal modes for gravitational and test fields do not differ considerably.

The general covariant equations for massless scalar, Dirac and electromagnetic fields have the forms
\begin{equation}\label{KGg}
\frac{1}{\sqrt{-g}}\partial_\mu \left(\sqrt{-g}g^{\mu \nu}\partial_\nu\Phi\right)=0,
\end{equation}
\begin{equation}\label{dirac}
\gamma^{\alpha} \left( \frac{\partial}{\partial x^{\alpha}} - \Gamma_{\alpha} \right) \Psi=0,
\end{equation}
\begin{equation}\label{EmagEq}
\frac{1}{\sqrt{-g}}\partial_\mu \left(F_{\rho\sigma}g^{\rho \nu}g^{\sigma \mu}\sqrt{-g}\right)=0.
\end{equation}
Here $F_{\rho\sigma}=\partial_\rho A_{\sigma}-\partial_\sigma A_{\rho}$, $A_\mu$ is a vector potential, $\gamma^{\alpha}$ are noncommutative gamma matrices and $\Gamma_{\alpha}$ are spin connections in the tetrad formalism.
After separation of variables, Eqs. (\ref{KGg}), (\ref{dirac}), (\ref{EmagEq}) can be reduced to the second order differential wavelike equation,
\begin{equation}\label{wave-equation}
\frac{d^2\Psi_s}{dr_*^2}+\left(\omega^2-V_{s}(r)\right)\Psi_s=0,
\end{equation}
where $s=0$ corresponds to the scalar field, $s=1/2$ to the Dirac field and $s=1$ to the electromagnetic field. The ``tortoise coordinate'' $r_*$ is defined by the relation
$$ dr_*=\frac{dr}{f(r)},$$ and the effective potentials are
\begin{equation}\label{scalarpotential}
V_{0}(r) = f(r)\left(\frac{\ell(\ell+1)}{r^2}+\frac{1}{r}\frac{d f(r)}{dr}\right),
\end{equation}
\begin{equation}
V_{\pm1/2}(r) = \frac{\ell+\frac{1}{2}}{r}\left(\frac{f(r) (\ell+\frac{1}{2})}{r}\mp\frac{\sqrt{f(r)}}{r}\pm\frac{d \sqrt{f(r)}}{dr}\right),
\end{equation}
\begin{equation}\label{empotential}
V_{1}(r) = f(r)\frac{\ell(\ell+1)}{r^2}.
\end{equation}
The effective potentials for scalar and electromagnetic fields have the form of a positive definite potential barrier with a single maximum.
The effective potential for the Dirac field with the minus sign in front of the derivative of $f(r)$ has negative gap near the event horizon. However, the potential with opposite chirality is positive definite and according to \cite{Zinhailo:2019rwd} the stability immediately follows for spherically symmetric black holes due to the isospectrality of both effective potentials.

Quasinormal modes $\omega_{n}$ ($n$ is the overtone number) correspond to solutions of the master wave equation (\ref{wave-equation}) with the requirement of the purely outgoing waves at infinity and purely incoming waves at the event horizon:
\begin{equation}
\Psi_{s} \sim \pm e^{\pm i \omega r^{*}}, \quad r^{*} \rightarrow \pm \infty.
\end{equation}

\begin{table}
\begin{tabular}{p{1.4cm}cccc}
\hline
$\lambda$ & Sixth order WKB ($\tilde{m} =5$) & Time domain  \\
\hline
0.1         & $0.109907-0.103986 i$    &  $0.110381-0.106662 i$  \\
5.1         & $0.098043-0.094181 i$    &  $0.096954-0.094142 i$  \\
10.1        & $0.093432-0.087633 i$    &  $0.090291-0.089237 i$  \\
15.1        & $0.089158-0.083783 i$    &  $0.086726-0.086373 i$  \\
20.1        & $0.085870-0.081412 i$    &  $0.084032-0.084297 i$  \\
25.1        & $0.083325-0.079730 i$    &  $0.081824-0.082673 i$  \\
30.1        & $0.081282-0.078419 i$    &  $0.080154-0.081257 i$  \\
35.1        & $0.079586-0.077339 i$    &  $0.078650-0.080080 i$ \\
40.1        & $0.078144-0.076416 i$    &  $0.077387-0.078936 i$  \\
45.1        & $0.076891-0.075608 i$    &  $0.076269-0.078012 i$  \\
49.6        & $0.075892-0.074956 i$    &  $0.075326-0.077278 i$  \\
\hline
\end{tabular}
\caption{The fundamental quasinormal mode of the scalar field ($\ell=0$, $n=0$, $M =1$) as a function of $\lambda$. }\label{tab1}
\end{table}

\begin{table}
\begin{tabular}{p{1.4cm}cccc}
\hline
$\lambda$ & Sixth order WKB ($\tilde{m} =5$) & Time domain  \\
\hline
0.1         & $0.181420-0.096074 i$    &  $0.181519-0.096383 i$  \\
5.1         & $0.153629-0.084293 i$    &  $0.153441-0.085286 i$  \\
10.1        & $0.142867-0.079712 i$    &  $0.142915-0.081136 i$  \\
15.1        & $0.136024-0.077057 i$    &  $0.136435-0.078456 i$  \\
20.1        & $0.131134-0.075223 i$    &  $0.131714-0.076507 i$  \\
25.1        & $0.127369-0.073806 i$    &  $0.128046-0.074954 i$  \\
30.1        & $0.124324-0.072642 i$    &  $0.125025-0.073694 i$  \\
35.1        & $0.121773-0.071650 i$    &  $0.121744-0.073107 i$ \\
40.1        & $0.119583-0.070786 i$    &  $0.120082-0.071668 i$  \\
45.1        & $0.117667-0.070018 i$    &  $0.118185-0.070931 i$  \\
49.6        & $0.116128-0.069394 i$    &  $0.116588-0.070183 i$  \\
\hline
\end{tabular}
\caption{The fundamental quasinormal mode of the Dirac field ($\ell=1/2$, $n=0$, $M =1$) as a function of $\lambda$. }\label{tab2}
\end{table}

\begin{table}
\begin{tabular}{p{1.4cm}cccc}
\hline
$\lambda$ & Sixth order WKB ($\tilde{m} =5$) & Time domain  \\
\hline
0.1         & $0.246431-0.091973 i$    &  $0.246416-0.092013 i$  \\
5.1         & $0.210903-0.081067 i$    &  $0.210776-0.081191 i$  \\
10.1        & $0.197552-0.076924 i$    &  $0.197448-0.077013 i$  \\
15.1        & $0.189197-0.074298 i$    &  $0.189093-0.074385 i$  \\
20.1        & $0.183132-0.072372 i$    &  $0.181134-0.075830 i$  \\
25.1        & $0.178386-0.070850 i$    &  $0.178269-0.070923 i$  \\
30.1        & $0.174496-0.069594 i$    &  $0.174529-0.069749 i$  \\
35.1        & $0.171208-0.068525 i$    &  $0.171089-0.068585 i$  \\
40.1        & $0.168364-0.067595 i$    &  $0.168279-0.067650 i$  \\
45.1        & $0.165861-0.066772 i$    &  $0.165754-0.066772 i$  \\
49.6        & $0.163842-0.066105 i$    &  $0.163741-0.066129 i$  \\
\hline
\end{tabular}
\caption{The fundamental quasinormal mode of the electromagnetic field ($\ell=1$, $n=0$, $M =1$) as a function of $\lambda$. }\label{tab3}
\end{table}

\begin{figure}
\centerline{\resizebox{\linewidth}{!}{\includegraphics*{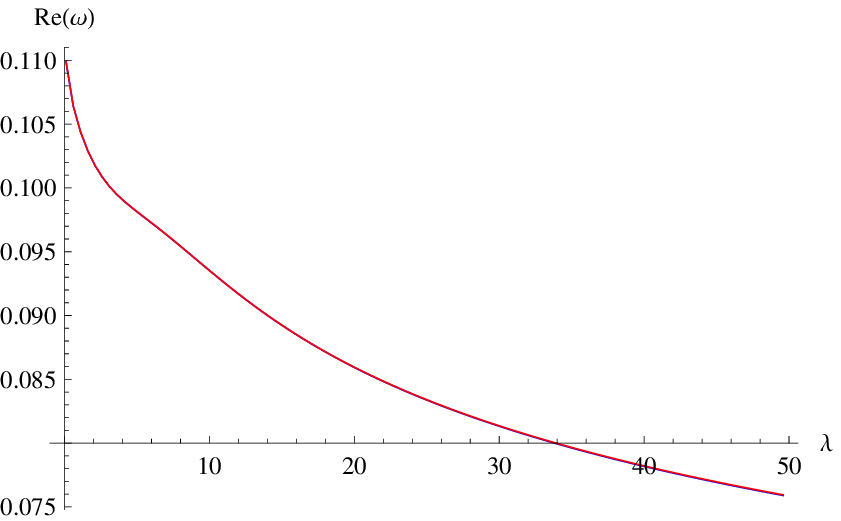}}}
\centerline{\resizebox{\linewidth}{!}{\includegraphics*{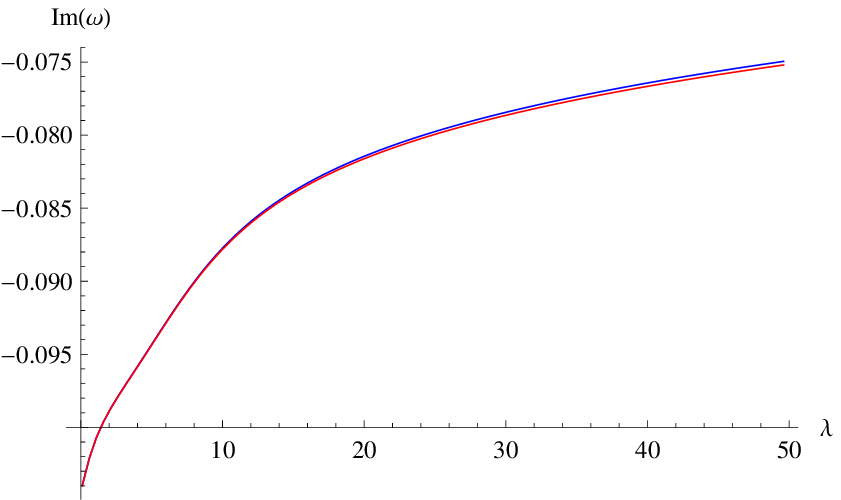}}}
\caption{The fundamental ($n=0$) quasinormal mode computed by the  sixth order WKB approach ($\tilde{m} =5$) for $\ell=0$ scalar perturbations as a function of $\lambda$, $M =1$, the blue line corresponds to the first order approximation ($b_1 \neq 0$, $b_2 =b_3 =...=0$): the red line corresponds to the second order approximation for the metric when $b_1 \neq 0$ and $b_2 \neq 0$.}\label{fig1}
\end{figure}

\begin{figure}
\centerline{\resizebox{\linewidth}{!}{\includegraphics*{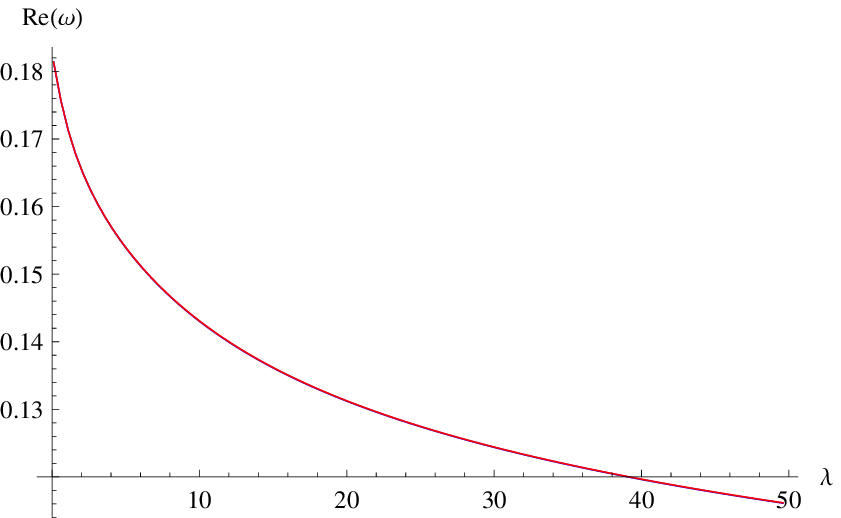}}}
\centerline{\resizebox{\linewidth}{!}{\includegraphics*{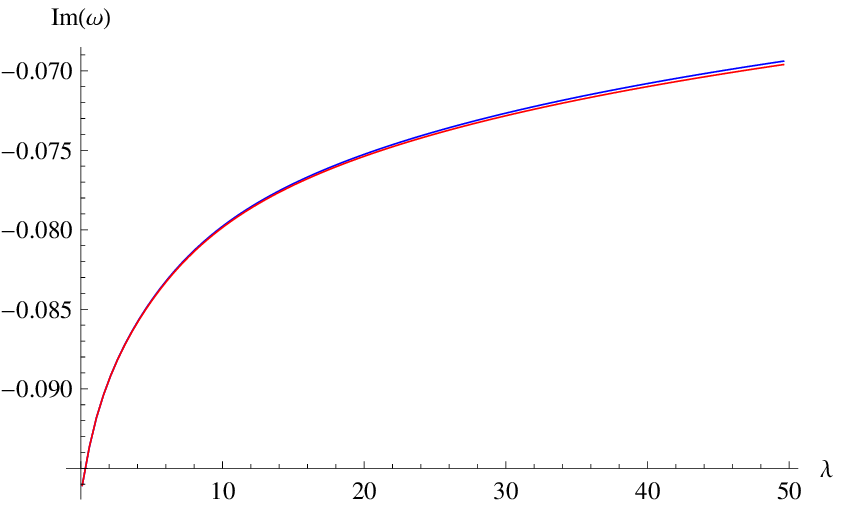}}}
\caption{The fundamental ($n=0$) quasinormal mode computed by the sixth order WKB approach ($\tilde{m} =5$) for $\ell=1/2$ Dirac perturbations as a function of $\lambda$, $M =1$, the blue line corresponds to the first order approximation  ($b_1 \neq 0$, $b_2 =b_3 =...=0$): the red line corresponds to the second order approximation for the metric when $b_1 \neq 0$ and $b_2 \neq 0$.}\label{fig2}
\end{figure}

\begin{figure}
\centerline{\resizebox{\linewidth}{!}{\includegraphics*{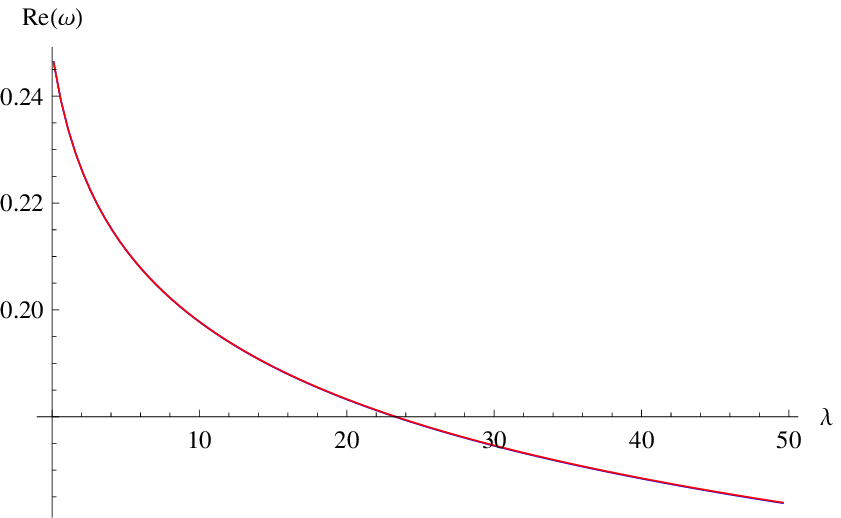}}}
\centerline{\resizebox{\linewidth}{!}{\includegraphics*{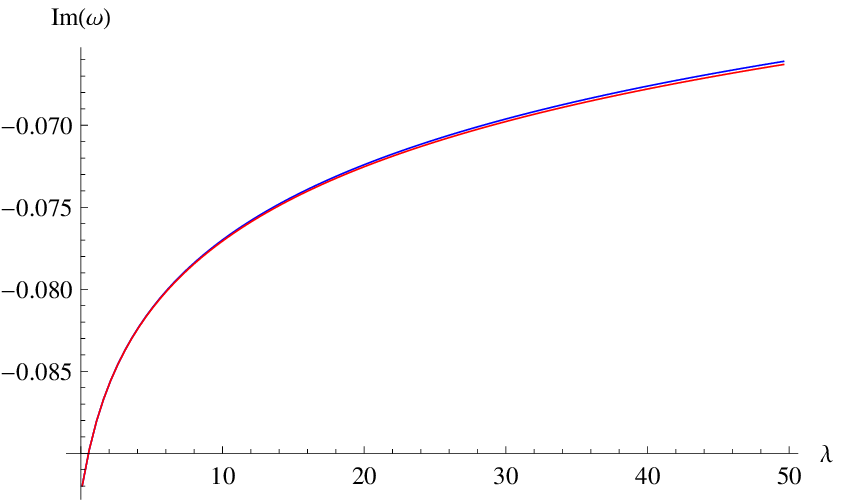}}}
\caption{The fundamental ($n=0$) quasinormal mode computed by the sixth order WKB approach ($\tilde{m} =5$) for $\ell=1$ electromagnetic perturbations as a function of $\lambda$, $M =1$, the blue line corresponds to the first order approximation  ($b_1 \neq 0$, $b_2 =b_3 =...=0$): the red line corresponds to the second order approximation for the metric when $b_1 \neq 0$ and $b_2 \neq 0$.}\label{fig3}
\end{figure}

\begin{figure}
\resizebox{\linewidth}{!}{\includegraphics*{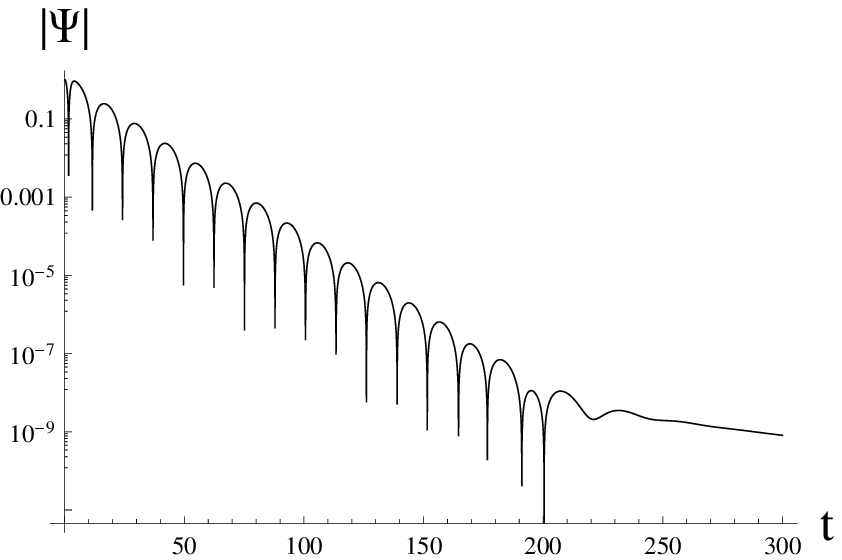}}
\caption{Time-domain profile for the electromagnetic field for the multipole number $\ell=1$, $\lambda=0.1$ in the units $M=1$.}\label{fig:TD}
\end{figure}

For finding of the low-laying quasinormal modes we will use the two independent methods:
\begin{enumerate}
\item Integration of the wave equation (before the introduction of the stationary ansatz, that is, keeping the second derivative in time instead of $\omega^2$ in the wave equation) in time domain at a given point in space \cite{Gundlach:1993tp}. We will integrate the wavelike equation rewritten in terms of the light cone variables $u=t-r_*$ and $v=t+r_*$. The appropriate discretization scheme was proposed in \cite{Gundlach:1993tp},
$$
\Psi\left(N\right)=\Psi\left(W\right)+\Psi\left(E\right)-\Psi\left(S\right)-
$$
\begin{equation}\label{Discretization}
-\Delta^2\frac{V\left(W\right)\Psi\left(W\right)+V\left(E\right)\Psi\left(E\right)}{8}+{\cal O}\left(\Delta^4\right)\,,
\end{equation}
where we used the following notation for the points:
$N=\left(u+\Delta,v+\Delta\right)$, $W=\left(u+\Delta,v\right)$, $E=\left(u,v+\Delta\right)$ and $S=\left(u,v\right)$. The initial data are given on the null surfaces $u=u_0$ and $v=v_0$. This method was used in a great number of works (see for example \cite{Konoplya:2008au,Churilova:2020bql,Konoplya:2019xmn,Konoplya:2014lha,Zhidenko:2008fp,Turimov:2019afv,Lin:2019fte,Dias:2020ncd} and references therein) and proved its efficiency for testing (in)stability \cite{Konoplya:2008au,Konoplya:2014lha,Dias:2020ncd}, because it takes into consideration contribution of all overtones for a given multipole number $\ell$.

\item In the frequency domain  we will use the WKB method of Will and Schutz \cite{Schutz:1985zz}, which was extended to higher orders in \cite{Iyer:1986np,Konoplya:2003ii,Matyjasek:2017psv} and made even more accurate by the usage of the  Padé  approximants in \cite{Matyjasek:2017psv,Hatsuda:2019eoj}.
The higher-order WKB formula has the form \cite{Konoplya:2019hlu},
$$ \omega^2=V_0+A_2(\K^2)+A_4(\K^2)+A_6(\K^2)+\ldots - $$
\begin{equation}\nonumber
\imo \K\sqrt{-2V_2}\left(1+A_3(\K^2)+A_5(\K^2)+A_7(\K^2)\ldots\right),
\end{equation}
where $\K$ takes half-integer values. The corrections $A_k(\K^2)$ of the order $k$ to the eikonal formula are polynomials of $\K^2$ with rational coefficients and depend on the values of higher derivatives of the potential $V(r)$ in its maximum. In order to increase accuracy of the WKB formula, we follow Matyjasek and Opala \cite{Matyjasek:2017psv} and use the Padé approximants. Here we will use the sixth order WKB method with $\tilde{m} =5$, where $\tilde{m}$ is defined in \cite{Matyjasek:2017psv,Konoplya:2019hlu}, because this choice provides the best accuracy for the Schwarzschild limit.
\end{enumerate}

Since both methods (the WKB method and time-domain integration) are very well known (see reviews \cite{Konoplya:2019hlu,Konoplya:2011qq}), we will not describe them in this paper in more detail, but will simply show that both methods are in a very good agreement in the common range of applicability.

The first question which we would like to respond is how much quasinormal modes for the first order approximation of the metric (that is, when $b_2 = b_3 =...=0$) differ from those for the second ($b_3 =b_4...=0$) and higher orders. In other words, which order of the metric approximation is sufficient for description of the black hole geometry.
From Figs. \ref{fig1}-\ref{fig3} one can see that already the first order approximation which is provided by only two parameters $\varepsilon$ and $b_1$ provides sufficient accuracy: adding next correction changes the quasinormal modes by a small fraction of $1 \%$. This happens because the metric function changes relatively softly in the region near the black hole approaching the asymptotic regime relatively slowly. This class of black hole metrics was called in \cite{Konoplya:2020hyk} ``moderate'' and is very well approximated by only a few parameters.

We also observe that the damping rate given by the imaginary part of the quasinormal frequency is decreasing when the coupling constant $\lambda$ is increased, which means longer lived modes once the cubic correction is turned on. The real oscillation frequency is decreasing as well when $\lambda$ grows. The results obtained with the help of the WKB method although known to be sufficiently accurate when the  Padé summation is applied still need additional check, which was performed by the time-domain integration. The results represented in Tables I, II, III show that there is a very good agreement between the two methods and the difference between the results obtained by both methods is much smaller than the effect, that is, the deviation of the quasinormal frequency from its Schwarzschild value.  The typical time domain profile is shown on Fig. \ref{fig:TD} and it has the power-law tail in the end of the ringdown phase. Let us notice that the worst situation as to WKB accuracy and comparison with time-domain data is the scalar $\ell=0$
mode, for which, on the one hand, the WKB approach is less accurate than for $\ell>n$ modes, and, on the other hand, there are usually only a few damped oscillations in the signal before the domination of the asymptotic power-law tails. Here, fortunately, power-law tails begin at sufficiently late times and several oscillations occurs even for the lowest $\ell=0$ multipole, so that the Prony method allows one to extract the value of the quasinormal frequency with the sufficient accuracy. Prolonged period of quasinormal ringing which we observe in the time domain is phenomenon that may depend on the initial wave packet rather than on the gravitational theory. Unlike quasinormal frequencies, this characteristic depends not only on the parameters of the black holes, but also on the initial conditions. Therefore, apparently, looking at different initial conditions mimicking real astrophysical processes, we could learn whether the prolonged period of quasinormal oscillations is an objective fact and not an artifact of the integration scheme and initial conditions.

\section{Grey-body factors}

\begin{figure}
\resizebox{\linewidth}{!}{\includegraphics*{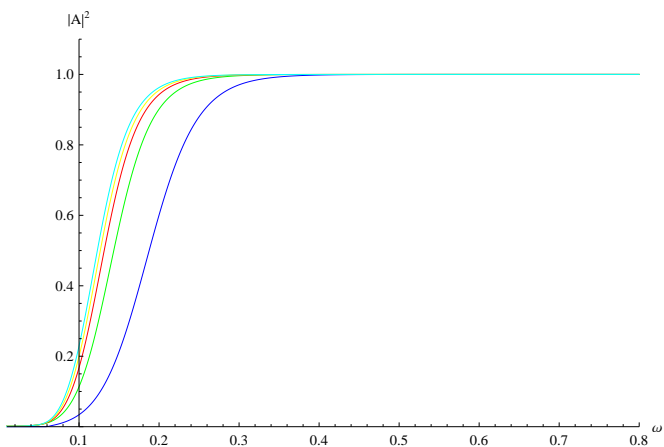}}
\caption{Grey-body factors for the Dirac field $k=1$, $\lambda=0.01$ (blue), $15$ (green), $30$ (red), $40$ (yellow), $50$ (light blue), $M=1$.}\label{fig:GBDirac}
\end{figure}
\begin{figure}
\resizebox{\linewidth}{!}{\includegraphics*{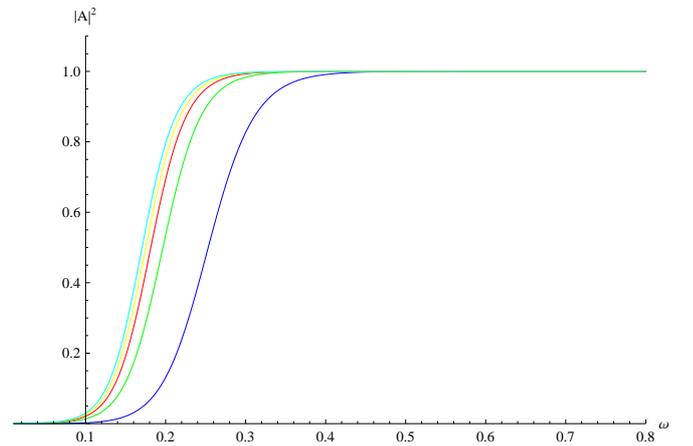}}
\caption{Grey-body factors for the Maxwell field $\ell=1$, $\lambda=0.01$ (blue), $15$ (green), $30$ (red), $40$ (yellow), $50$ (light blue), $M=1$.}\label{fig:GBMaxwell}
\end{figure}

In order to calculate the fraction of particles scattered back by the effective potential to the event horizon and learn which is the flow of particles which reaches the distant observer, we need to solve the spectral problem with different (from quasinormal) boundary conditions.
We will study wave equation (\ref{wave-equation}) with the boundary conditions allowing for incoming waves from infinity. Owing to the symmetry of the scattering properties this is identical to the scattering of a wave coming from the horizon, which is natural if one wants to know the fraction of particles reflected back to the horizon. Thus, the scattering boundary conditions for Eq. (\ref{wave-equation}) have the form
\begin{equation}\label{BC}
\begin{array}{ccll}
    \Psi_{\ell} &=& e^{-i\omega r_*} + R_{\ell} e^{i\omega r_*},& r_* \rightarrow +\infty, \\
    \Psi_{\ell} &=& T_{\ell} e^{-i\omega r_*},& r_* \rightarrow -\infty, \\
\end{array}%
\end{equation}
where $R_{\ell}$ and $T_{\ell}$ are called the reflection and transmission coefficients (for a given multipole number $\ell$), so that one has
\begin{equation}\label{1}
\left|T_{\ell}\right|^2 + \left|R_{\ell}\right|^2 = 1.
\end{equation}
Once the reflection coefficient is found, we can calculate the transmission coefficient for each $\ell$ using the WKB approach as follows:
\begin{equation}
\left|A_{\ell}\right|^2=1-\left|R_{\ell}\right|^2=\left|T_{\ell}\right|^2.
\end{equation}
\begin{equation}\label{moderate-omega-wkb}
R = (1 + e^{- 2 i \pi K})^{-\frac{1}{2}},
\end{equation}
where $K$ can be found  from the following equation:
\begin{equation}
K - i \frac{(\omega^2 - V_{0})}{\sqrt{-2 V_{0}^{\prime \prime}}} - \sum_{i=2}^{i=6} \Lambda_{i}(K) =0.
\end{equation}
Here $V_0$ is the maximum of the effective potential, $V_{0}^{\prime \prime}$ is the second derivative of the
effective potential in its maximum with respect to the tortoise coordinate $r_{*}$, and $\Lambda_i$  are higher order WKB corrections which depend on up to $2i$th order derivatives of the effective potential at its maximum \cite{Schutz:1985zz,Iyer:1986np,Konoplya:2003ii,Matyjasek:2017psv,Hatsuda:2019eoj} and $K$. This approach at the sixth WKB order was used for finding transmission/reflection coefficients of various black holes and wormholes in \cite{Konoplya:2019ppy,Konoplya:2019hml}, and the comparison of the WKB results for the energy emission rate of Schwarzschild black hole done in \cite{Konoplya:2019ppy} is in an excellent concordance with the numerical calculations of the well-known work by Don Page \cite{Page:1976df}. Here we will mostly use the sixth order WKB formula of \cite{Konoplya:2003ii} and, sometimes, apply lower orders when small frequencies and lower multipoles are under consideration. Fortunately, the WKB method works badly for small frequencies only, that is, in the region where the reflection is almost total and the grey-body factors are close to zero. Therefore, this inaccuracy of the WKB approach at small frequencies does not affect our estimations of the energy emission rates.

From Figs. \ref{fig:GBDirac}, \ref{fig:GBMaxwell} one can see that the grey-body factors for a given value of the real frequency $\omega$ for both Dirac and Maxwell fields are considerably increased when the coupling constant $\lambda$ is turned on. This means that the height of the potential barrier surrounding the black hole is smaller at the increasing $\lambda$, which allows a greater number of particles to penetrate the barrier. Thus, the grey-body factors work for the enhancing of the Hawking radiation. However, the total effect usually depends more on the temperature of the black hole than on the grey-body factors, and this aspect will be studied in the next section, where we will calculate the corresponding energy emission rates.

\begin{figure}
\resizebox{\linewidth}{!}{\includegraphics*{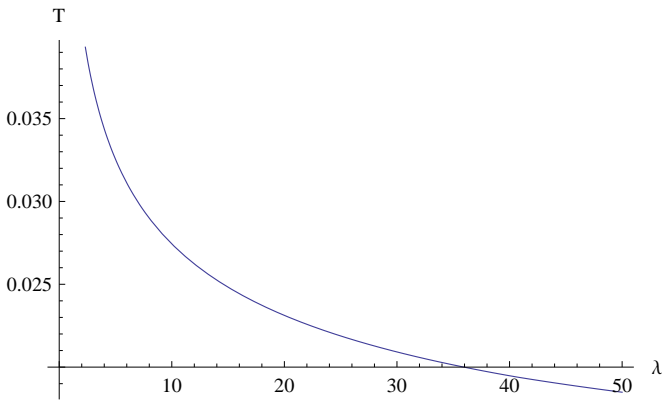}}
\caption{Hawking temperature as a function of $\lambda$, $M=1$.}\label{figTemperature}
\end{figure}
\begin{figure*}
\resizebox{\linewidth}{!}{\includegraphics*{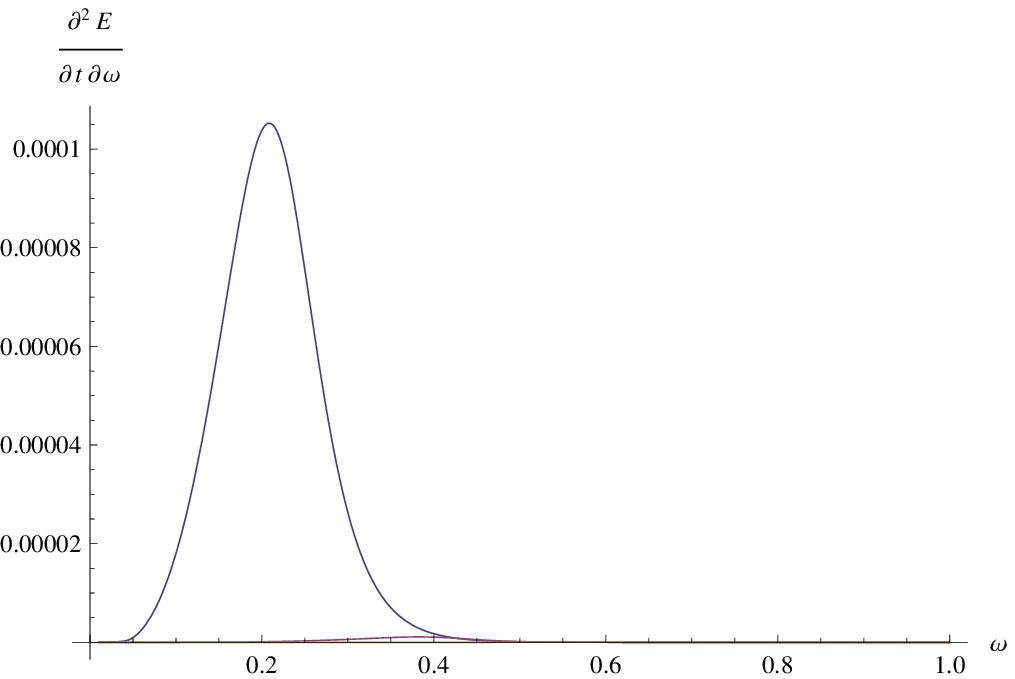}\includegraphics*{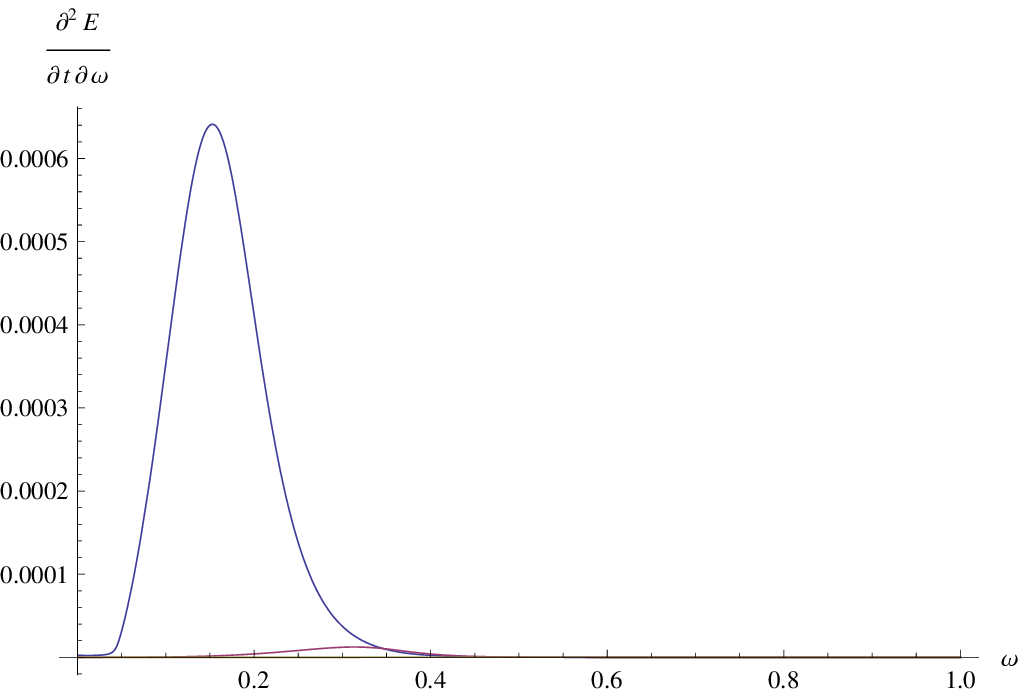}}
\caption{Energy emission rate $\partial^{2} E/\partial t \partial \omega $ for the Maxwell (left) and Dirac (right) fields as a function of $\omega$, $M=1$, $\lambda=2.5$. Blue is for $\ell=1$ ($k=1$ for Dirac); red is for $\ell=2$ ($k=2$ for Dirac). The contribution of the third multipole is very small, but still used in the calculations.  }\label{figDistribution}
\end{figure*}
\begin{figure*}
\resizebox{\linewidth}{!}{\includegraphics*{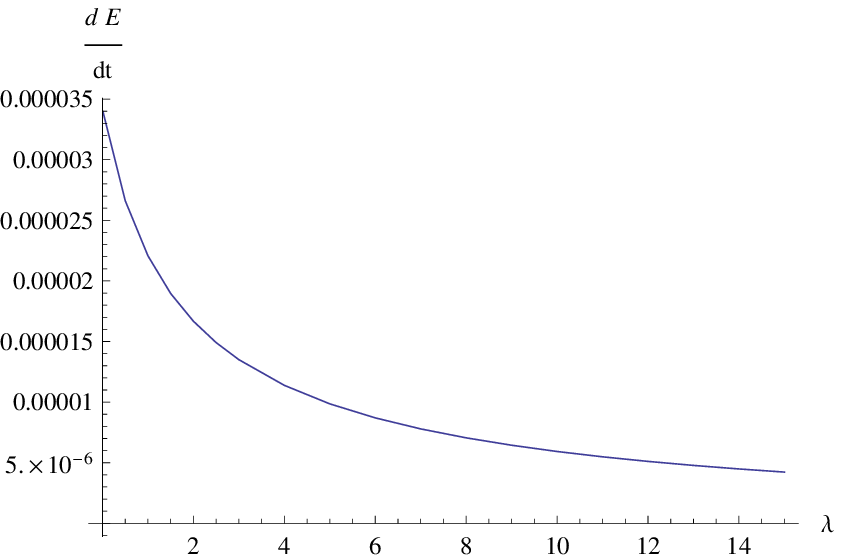}\includegraphics*{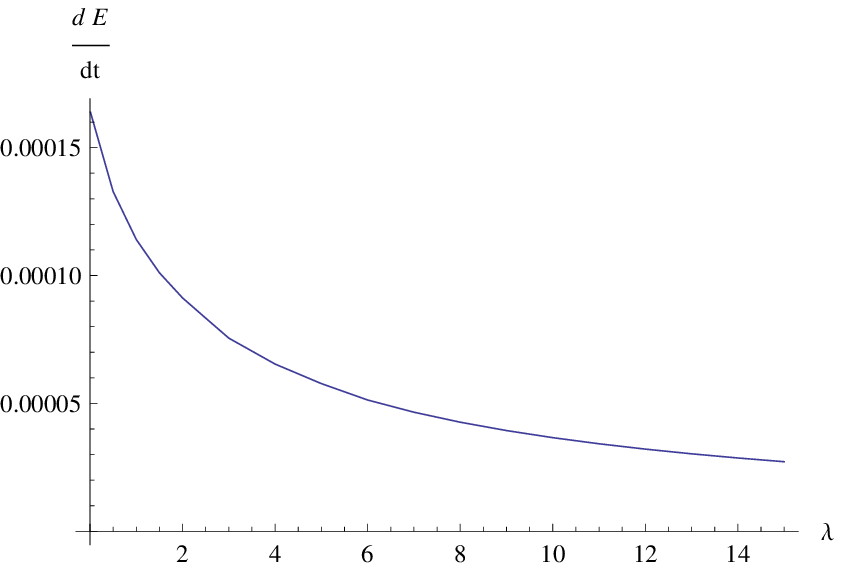}}
\caption{Total emission $dE/dt$ for the Maxwell (left) and Dirac (right) fields as a function of $\lambda$, $M=1$.}\label{fig:TotalDirac}
\end{figure*}
\begin{figure}
\resizebox{\linewidth}{!}{\includegraphics*{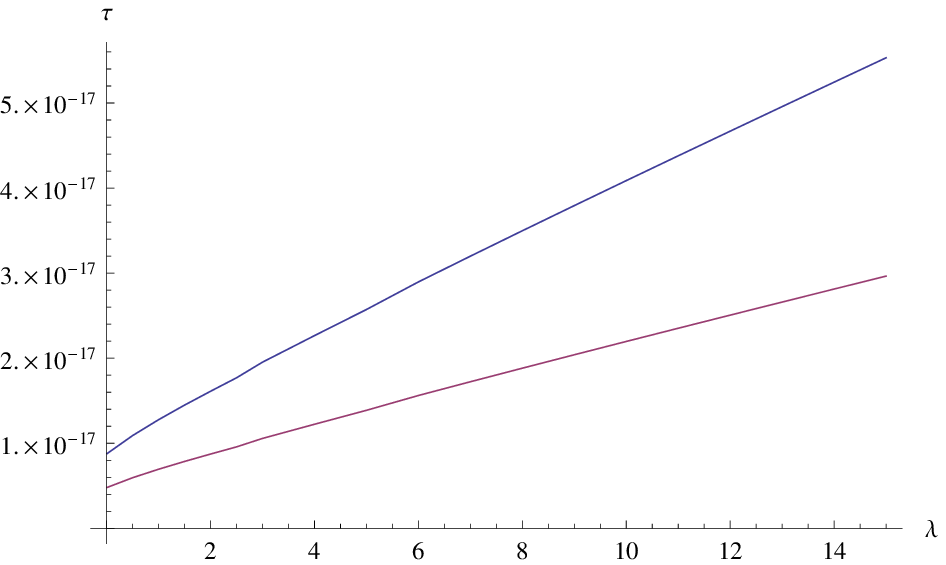}}
\caption{Lifetime of the black hole $\tau$ as a function of $\lambda$, $M=1$ in the usual (top) and ultrarelativistic (bottom) regimes.}\label{figLifeTime}
\end{figure}

\section{Intensity of Hawking radiation and the black hole lifetime}

We will assume that the black hole is in the thermal equilibrium with its environment in the following sense: the temperature of the black hole does not change between the emissions of two consequent particles. This implies that the system can be described by the canonical ensemble \cite{Hawking:1974sw}. Therefore, the energy emission rate for Hawking radiation is calculated by the formula \cite{Hawking:1974sw},
\begin{equation}\label{energy-emission-rate}
\frac{\text{d}E}{\text{d} t} = \sum_{\ell} N_{\ell} \left| A_{\ell} \right|^2 \frac{\omega}{\exp\left(\omega/T \right)\pm1} \frac{\text{d} \omega}{2 \pi},
\end{equation}
were $T_H$ is the Hawking temperature, $A_{\ell}$ are the grey-body factors, and $N_{\ell}$ are the multiplicities, which only depend on the space-time dimension and $\ell$. The Hawking temperature for spherically symmetric black hole is
\begin{equation}
T = \frac{1}{4 \pi} f'(r) \bigg|_{r=r_{0}}.
\end{equation}
The multiplicity factors for the four-dimensional spherically symmetrical black holes case consists of
the number of degenerated $m$-modes (which are $m = -\ell, -\ell+1, ....-1, 0, 1, ...\ell$, that is  $2 \ell +1$ modes) multiplied by the number of species of particles which depends also on the number of polarizations and helicities of particles. Therefore, we have
\begin{equation}
N_{\ell} = 2 (2 \ell+1) \qquad (Maxwell),
\end{equation}
\begin{equation}
N_{\ell} = 8 k \qquad (Dirac).
\end{equation}
Here $k=\ell + 1/2$ for the Dirac field. The multiplicity factor for the Dirac field is fixed taking into account both the ``plus'' and ``minus'' potentials which are related by the Darboux transformations, which leads to the isospectral problem and the same grey-body factors for both chiralities. We will use here the ``minus'' potential, because the WKB results are more accurate for that case in the Schwarzschild limit.

From Fig. \ref{figTemperature} one can see that Hawking temperature monotonically decays when the coupling constant $\lambda$ is increased. Therefore, the temperature factor, unlike transmission coefficients calculated in the previous section, works for suppression of the Hawking radiation. We can see that, as a result, the exponential temperature factor becomes more influential: the total energy emission rates  of both electromagnetic and Dirac fields are monotonically decreasing (see Fig. \ref{fig:TotalDirac}). Notice that also intuitively it would be expected that the colder black hole will provide smaller flux of radiation to a distant observer, this is not always so, and for example, in the Einstein-Weyl theory \cite{Konoplya:2019ppy}.

There are two different regimes of emission of particles  \cite{Page:1976df}: a) when the black hole mass is sufficiently large, so that the radiation of massive particles can be neglected and the flux consists mainly of massless electron and muon neutrinos, photons, and gravitons and b) when the black hole mass is sufficiently small and emission of electrons and positrons will occur ultrarelativistically. In the second case, the  radiation of electrons and positrons can be approximated by the massless Dirac field and the emission rate of all the Dirac particles must be  doubled. Supposing that the peak in the Dirac particles' spectrum  $\partial^{2} E/\partial t \partial \omega$ occurs at some $\omega \approx \xi M^{-1} $, we can see from fig. \ref{figDistribution} that this peak almost does not change when $\lambda$ is increased. The same is true even for sufficiently large $\lambda$. Therefore, the range of energies in which the ultra-relativistic regime of radiation takes place is roughly the same in the Einsteinian cubic gravity as for the Schwarzschild black hole, that is, at
\begin{equation}\nonumber
m_{e} = 4.19 \times 10^{-23} m_{p} \ll \xi M^{-1} \ll m_{\mu} = 8.65 \times 10^{-21} m_{p}.
\end{equation}

The energy emitted causes the black hole mass to decrease at the following rate \cite{Page:1976df}
\begin{equation}
\frac{d M}{d t} = -\frac{\hbar c^4}{G^2} \frac{\alpha_{0}}{M^2},
\end{equation}
where we have restored the dimensional constants. Here $\alpha_{0} = d E/d t$ is taken for a given initial mass $M_{0}$. Since most of its time the black hole spends near  its original state $M_{0}$ and integrating of the above equation  gives us the lifetime of a black hole,
\begin{equation}
\tau = \frac{G^2}{\hbar c^4} \frac{M_{0}^3}{3 \alpha_{0}}.
\end{equation}

From Fig. \ref{figLifeTime} one can see that the lifetime of the black hole is increased by almost one order in comparison with the Schwarzschild limit (for which we reproduce the results of \cite{Page:1976df}). The ultrarelativistic emission is characterized by more intensive evaporation process (lower line). At large values of the coupling constant $\lambda$ the lifetime $\tau$ is roughly proportional to $\lambda$,
\begin{equation}
\tau \approx 8.7\times 10^{-18} (1+ 0.36 \lambda),
\end{equation}
and for the ultrarelativistic regime we have
\begin{equation}
\tau \approx 4.8\times 10^{-18} (1+ 0.36 \lambda).
\end{equation}

Here we did not consider emission of gravitons. However, as is known from a number of papers, in the four-dimensional theories contribution of gravitons in the total energy emission is usually very small. It consists of about $1\% -2\%$  of the total emission in the Schwarzschild black hole \cite{Page:1976df} and in the 4D Einstein-Gauss-Bonnet black holes \cite{Konoplya:2020cbv}. As the effect, that is the deviation of the energy emission rate from its Schwarzschild value, exceeds $100 \%$, here we can safely neglect contribution of gravitons for qualitative understanding of the Hawking radiation.

Thermodynamic properties of a more general class of black holes with higher curvature  correction have been recently studied in \cite{Bueno:2017qce}.
There is not any data which could be used for the direct comparison with our results, because here we concentrate on the cubic theory and calculations of the intensity of Hawking radiation and grey-body factors. Nevertheless, there are two important conclusions made in  \cite{Bueno:2017qce} which also supports our results. First is about the duration of the semiclassical regime. In Table I of \cite{Bueno:2017qce} it was noticed that the breakdown of the semiclassical regime occurs at a somewhat different minimal mass which depends on the new energy scale and this means that the semiclassical regime is clearly determined and the appropriate standard approaches can be used for calculations of intensity of Hawking radiation. In addition, it is noticed in \cite{Bueno:2017qce} that the temperature of the higher curvature corrected black hole is usually smaller, and the estimation of the order of the evaporation time  (without taking into consideration the grey-body factors) shows that the lifetime is many orders higher when the higher curvature corrections are tuned on.
This qualitatively agrees with our calculations which were limited by moderate values of the coupling constant and, nevertheless, detected strong suppression of Hawking radiation.

\section{Discussion}

In this work for the first time we have calculated quasinormal modes of the scalar, Dirac and electromagnetic fields in the background of the four-dimensional black hole in the Einsteinian cubic gravity. We also computed the grey-body factors for fields representing emission of photons, electrons, positrons and neutrinos. We have shown that:

\begin{itemize}
\item When the coupling constant $\lambda$ representing cubic correction to the Einstein term is increasing, the damping rate and the real oscillation frequencies are suppressed.
\item The grey-body factors are larger for nonzero values of $\lambda$, which works for increasing the amount of radiation that will reach the observer.
\item Despite such behavior of the grey-body factors, the temperature is falling when $\lambda$ is tuned on and the total energy emission rate for all the considered types of particles is decreased, which leads to the slower evaporation of the black hole.
\item At moderate and large values of the coupling constant $\lambda$ the lifetime of the black hole is roughly proportional to $\lambda$.
\end{itemize}

There are a number of open questions which were beyond the scope of this publication. First of all, this concerns gravitational perturbations, which are important not only for estimating the constrains on higher curvature corrections from the observation of gravitational waves \cite{Blazquez-Salcedo:2016enn,Ayzenberg:2013wua,deRham:2020ejn} but also because they allow us to test the stability of the black hole \cite{Konoplya:2017lhs,Takahashi:2010ye,Cuyubamba:2016cug,Takahashi:2012np,Konoplya:2020juj}. The stability region is essential when higher curvature corrections are included, as we know from the example of various quadratic theories of gravity.
Finally, the  slowly rotating black hole which was announced  recently in \cite{Adair:2020vso} deserves analysis of the quasinormal spectra and Hawking radiation. However, the separation of variables most probably will be impossible for this case.

\acknowledgments{
R. K. and A. Z. would like to thanks Robert Mann for useful discussions and Robie Hennigar for sharing his numerical data given on fig. \ref{fig:a} with us. This work was supported by 19-03950S GAČR grant and the ``RUDN University Program 5-100''. A. F. Z. acknowledges the SU grant SGS/12/2019 and 19-03950S GAČR grant.}

\end{document}